\documentclass[conference]{IEEEtran}
\IEEEoverridecommandlockouts
\usepackage{cite}
\usepackage{amsmath,amssymb,amsfonts}
\usepackage{algorithmic}
\usepackage{graphicx}
\usepackage{textcomp}
\usepackage{xcolor}
\usepackage{CJK}
 
\usepackage{booktabs}
\usepackage{algorithm,multirow,epstopdf,subfigure}
\usepackage{amsthm}
\theoremstyle{plain}

\newcommand{\methodshort}{SemiGNN}

\def\BibTeX{{\rm B\kern-.05em{\sc i\kern-.025em b}\kern-.08em
    T\kern-.1667em\lower.7ex\hbox{E}\kern-.125emX}}
\begin{document}
\begin{CJK}{UTF8}{gbsn}
\title{A Semi-supervised Graph Attentive Network for Financial Fraud Detection}

\author{Daixin Wang$^{1,2}$, Jianbin Lin$^1$, Peng Cui$^2$, Quanhui Jia$^1$, Zhen Wang$^1$, \\
        Yanming Fang$^1$, Quan Yu$^1$, Jun Zhou$^1$, Shuang Yang$^1$, Yuan Qi$^1$ \\
        $^1$Ant Financial Services Group, China \\
        $^2$Department of Computer Science and Technology, Tsinghua University, China
        }
\maketitle

\begin{abstract}
With the rapid growth of financial services, fraud detection has been a very important problem to guarantee a healthy environment for both users and providers. Conventional solutions for fraud detection mainly use some rule-based methods or distract some features manually to perform prediction. However, in financial services, users have rich interactions and they themselves always show multifaceted information. These data form a large multiview network, which is not fully exploited by conventional methods. Additionally, among the network, only very few of the users are labelled, which also poses a great challenge for only utilizing labeled data to achieve a satisfied performance on fraud detection. 

To address the problem, we expand the labeled data through their social relations to get the unlabeled data and propose a semi-supervised attentive graph neural network, named \methodshort{} to utilize the multi-view labeled and unlabeled data for fraud detection. Moreover, we propose a hierarchical attention mechanism to better correlate different neighbors and different views. Simultaneously, the attention mechanism can make the model interpretable and tell what are the important factors for the fraud and why the users are predicted as fraud. Experimentally, we conduct the prediction task on the users of Alipay, one of the largest third-party online and offline cashless payment platform serving more than 4 hundreds of million users in China. By utilizing the social relations and the user attributes, our method can achieve a better accuracy compared with the state-of-the-art methods on two tasks. Moreover, the interpretable results also give interesting intuitions regarding the tasks.
\end{abstract}


\section{Introduction}
Nowadays financial services especially online financial services provide much convenience to a lot of people and meanwhile create a huge economic benefit to the society. However, we are also witnessing more and more financial frauds. For example, in the US alone, the number of customers who experienced fraud hit a record $15.4$ million people, which is $16$ percent higher than 2015. Fraudsters stole about $6$ billion from banks last year\footnote{https://www.altexsoft.com/whitepapers/fraud-detection-how-machine-learning-systems-help-reveal-scams-in-fintech-healthcare-and-ecommerce/}. There are many kinds of fraud, like case-out fraud in credit-card services, insurance fraud, default and so on. These frauds all will seriously damage the security to both the users and the service providers. Therefore, how to do fraud detection is an important problem to be investigated. 

The goal of fraud detection is to predict whether an entity, which can be a user or a device or more, will be involved in a fraud or not in the future. Generally, the problem can be formulated as a classification problem. Conventional methods for fraud detection can be classified into two categories. The first kind is the rule-based methods. Their assumption is that financial fraudulent activities can be detected by looking at on-surface and evident signals. These signals can help design some rules to do fraud prediction. Although rule-based methods have served the industry for a long time, they have their own disadvantages. Rule designs are heavily relied on the human prior knowledge. Therefore, these methods are difficult to handle the changing and complex patterns. Furthermore, rule-based methods are easy to be attacked \cite{b1}. To address their limitations, machine learning methods are proposed to automatically mine the fraud patterns from data. Most of the machine learning methods extract user's statistical features from different aspects, such as user profile, user behaviors and transaction summarizing. These methods make prediction mainly based on the statistical features of a certain user and use classical classifiers like logistic regression, neural networks to do classification \cite{b2,b3,b4}. However, these methods seldom consider the interactions between users. Actually, there are rich interactions in the financial scenarios. For example, users have social relationships like friends, classmates and relatvies between each other. Users may have transactions with merchants or other users. Users have to login in some apps to achieve the financial transactions. All of these relationships may be beneficial to the fraud detection problem. Then some following methods start to use graph embedding to incorporate the user interactions \cite{b5,b6}. However, for the fraud detection, very few of the data are labeled and we usually have a large number of unlabeled data, which have not been fully exploited in existing graph-based methods. Furthermore, interpretable models and results are often preferred in financial scenarios but existing graph embedding methods are often black-box models. 

% Traditionally, occupation prediction was studied in the areas of psychology and economics \cite{b3,b4,b5}. They find the relationships between jobs and some personal profiles like gender, education and the languages the people use. Motivated by the conclusions of these works, some following works start to use machine learning technology to study this problem on larger datasets. Some methods mainly use the user-profile-related features to do the prediction and these methods do not consider any relations and interactions between users \cite{b6,b7}. However, a large proportion of the users are so inactive that they produce very few or even no content, which makes it impossible to solely utilize user generated content to infer user occupation. Then some works start to use the network structures to deal with this challenge \cite{b8,b9,b10}. They assume that people who belong to the same social circles often have similar occupations. 

Considering the limitations of existing methods, we aim to propose a method which can utilize both labelled and unlabelled multiview data to do fraud detection. However, it faces the following challenges: (1) \textbf{How to bridge the labeled data with the unlabeled data?}: Very few of the users are labeled as fraud or not. Therefore, only modelling the labeled data is difficult to obtain a satisfied performance. To incorporate the unlabeled data, how to bridge the relationships between the unsupervised information and the supervised information is the first challenge. (2) \textbf{How to model the data heterogeneity?}: Utilizing multiview data can provide a more comprehensive information regarding the task. Nevertheless, multiview data like social relations and user attributes have different statistical properites. Such a heterogeneity poses a great challenge to integrate multiview data. (3) \textbf{How to learn an interpretable model?} The results of fraud detection are often served for the financial risk control. No matter for the financial service providers, or for the supervisors, they all require interpretable models to have a better knowledge of the predicted results. Therefore, how to design an interpretable model for fraud detection is also challenging.

To address above three challenges, we propose a semi-supervised graph attentive neural model for multiview data, named \methodshort{}. The basic idea of \methodshort{} is to enhance the representations of users by fully exploiting the relational data and attribute data of both labeled and unlabled data. In detail, we bridge the labeled and unlabeled users via their social relations. And for each user, we also use his attributes to build the attribute network. The social relations and the attributes together form a large multiview network. Then we propose a semi-supervised graph neural network to simultaneously model the multiview information within network to do fraud detection. The model has several advantages: (1) It can fully exploit the supervised information and unsupervised structural information for fraud detection. (2) Our model can integrate multiview data to get a comprehensive result for fraud detection. (3) Building the attribute network instead of using attributes as the dense features can enhance the representation ability of the attribute information. Furthermore, we design a hierarchical attention mechanism into \methodshort{}. The first-layer node-level attention is designed to effectively correlate different neighbors or different attributes of users. The second-layer view-level attention is able to correlate different views of data. Moreover, the attention-based model can also provide interpretable results to give more insights regarding the task. 

In summary, the contributions of the paper are summarized as follows:
\begin{itemize}
	\item To the best of our knowledge, this work is among the first to introduce semi-supervised graph neural network for fraud detection problem. By utilizing both the labeled and unlabeled data, our model can extract the discriminative and structural information to get a more accurate classification result. 
	\item We propose a novel semi-supervised graph embedding model with hierarchical attention to model multiview graph for fraud detection and give interpretable results. The model is able to learn a sophisticated way to integrate the node's neighbors and differnet views of data.
	\item Experimental results on Alipay users in two tasks demonstrate that our proposed method achieves a substantial gain over state-of-the-art methods. Moreover, the interpretable results also give deep insights regarding the task. 
\end{itemize}

\section{Related Work}
\subsection{Financial Fraud Detection}
Empirically, our work is related to the problem of fraud detection. Financial fraud is an issue which has serious reaching consequences in both the finance industry and daily life. Therefore, a great number of literatures have been studied on different types of the fraud, like financial statement fraud \cite{b7,b8,b9}, credit card fraud \cite{b10,b11}, insurance fraud \cite{b12,b13} and so on. Earilier works mainly use the rule-based methods for fraud detection. They assume that the fraud activities have some obvious patterns. Accordingly, these works define some combinatorial rules to detect these fraud activities. Due to the simplicity and interpretability of rule-based methods, they are popular for fraud detection. However, rule-based methods are highly dependent on the human expert knowledge. They are difficult to find complex and changing patterns. And they are also easier to be attacked once the rules are awared by attackers. 

Considering the limitations of the rule-based methods, recent methods start to use the machine learning models to automatically find the intrinsic fraud patterns from data. Commonly, these methods first extract statistical features from different aspects like user's profiles and historical behaviours and then use some classical classifiers like SVM \cite{b9}, tree-based methods \cite{b10} or neural networks \cite{b7,b8} to decide fraud or not. For example, Ravisankar et al. \cite{b9} extracts some statistic features about assets, liabilities, incomes, debt and sales and uses a multi-layer feedforward network to do financial statement fraud detection. Kirkos et al. \cite{b8} selects some personal variables regarding the financial distress, debt structure, need for continuing growth, accounts receivable and totally uses a 27-dimensional vector as the input to the decision tree, neural network and bayesian belief network. They find that the bayesian belief networks outperforms the rest of the two methods. 

Aforementioned methods regard each entity as an individual and thus only consider personal attributes. However, in financial scenarios, entities have many interactions with each other, which will form a graph. Then a few of recent works start to utilize the graph for fraud detection. For example, Liu et al. \cite{b5} proposes a graph neural network for malicious account detection. Hu et al. \cite{b6} proposes a meta-path based graph embedding method for user cash-out prediction. They all demonstrate that the graph structure benefits the task of fraud detection a lot. However, as we stated before, their methods do not exploit the unlabeled data and are not interpretable. 

\subsection{Learning over graphs}
Technically, our work is related to the graph-based methods. Network embedding is an effective method to model the structure of a graph. It aims to learn a low-dimensional vector-representation for each node. Early works mainly focus on the pure network without node attributes. DeepWalk \cite{b19} and Node2vec \cite{b20} propose to use the random walk and skip-gram to learn the node representations. LINE \cite{b21} and SDNE \cite{b22} propose explicit objective functions for preserving first- and second-order proximity. Some further works \cite{b23,b24} use the matrix factorization to factorize high-order relation matrix. However, network data often come with the node and edge attributes. Then some further works like Metapath2vec \cite{b25} , HNE \cite{b26} are proposed to consider both the network topology and node features. Recently, graph convolution network based methods are very popular \cite{b27,b28}. They are inductive methods which can simulataneously learn with the network topology and node attributes. Unfortunately, these methods are usually designed for common tasks like link prediction and only exploit partial information in networks. Therefore they are suboptimal in terms of classificatiom performance and cannot provide interpretable results in the fraud detection problem.

Our work is also related to the graph-based semi-supervised learning (GSSL) method \cite{b29}. They treat labeled and unlabeled data as a vertex and learn a classifier which is consistent with the labels while making sure that the prediction results for similar vertex are also similar. Different GSSL algorithms apply different functions for graph regularization. However, these methods seldom consider the vertex's features and very few of these methods are applied to the multi-view and interpretable scenario. 

\section{The Model}

\subsection{Problem Definition and Notations}
\begin{figure}[htb]
\centering
\includegraphics[width=0.45\textwidth]{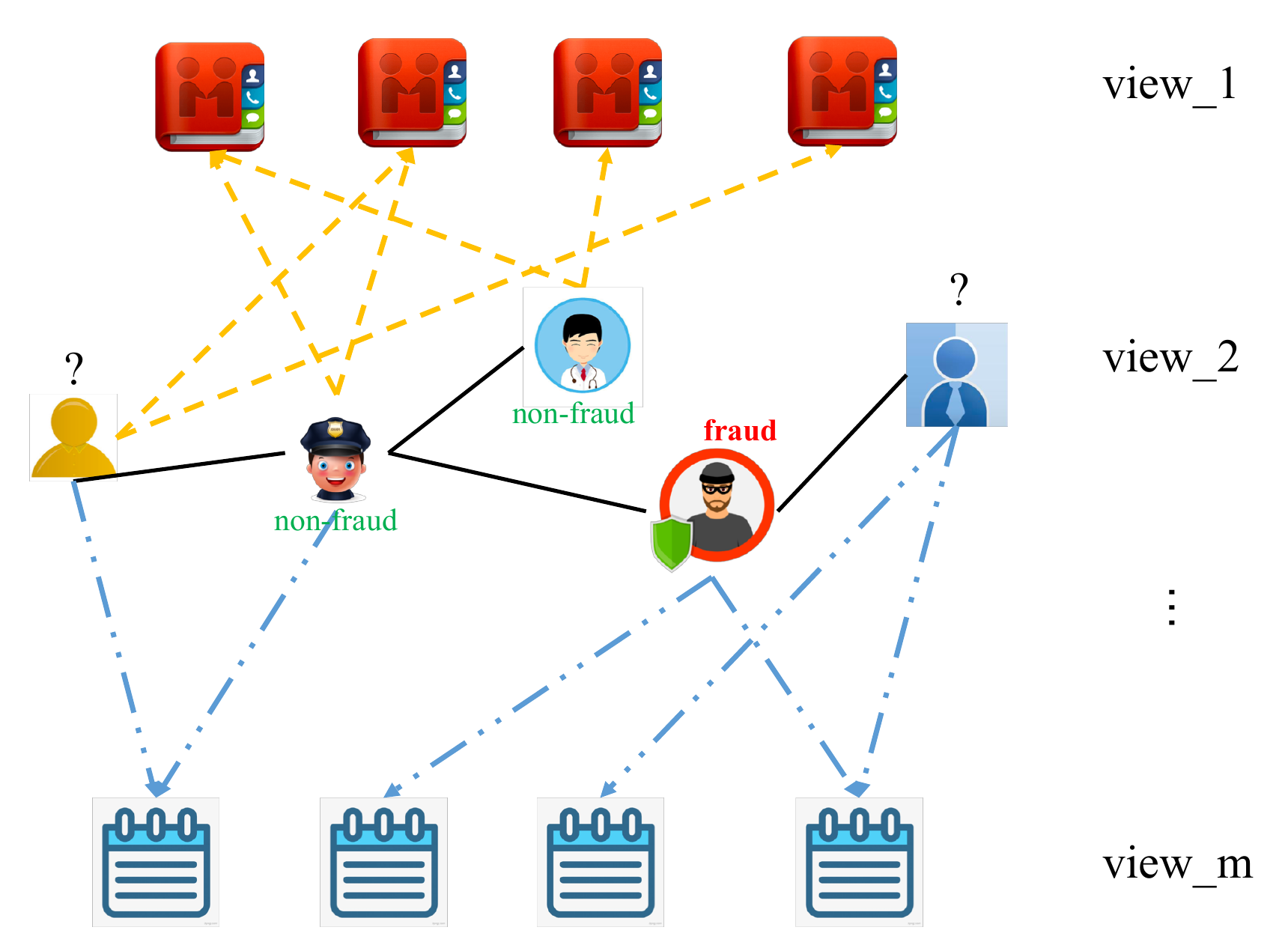}
\caption{The problem illustrations. Some users are labeled and most of the users are unlabeled. We collect different views of data. We aim to utilize the multiview data to do user fraud classification. }
\label{pic:data_framework}
\end{figure}

We first give a general description of our problem, as illustrated in Figure \ref{pic:data_framework}. In our problem, we collect multiple views of data, which denotes different facets of user information. Some views naturally form the graph, like social relations and transaction relations. But for some user attributes, we also formulate them as an attribute graph instead of dense features. We find that in this way our graph model can better find the correlations between the attributes. Note that here we propose a general model. But for different scenarios, we have different number of views and each view represents different information. We will specify the details of the dataset in the experiment. With the multiview network, our target is to train a classifier for user fraud classification. 

We summarize the notations of this paper in Table \ref{tab:notation}. We have $n_L$ labeled users denoted as $U_L$, each of which is labeled as fraud or not. From these users, we go through the social relations of the labeled users and thus get $n_{uL}$ unlabeled users, denoted as $U_{uL}$. We have $n_{uL} \gg n_L$. For each user, we have $m$ views of data. In each view, we have a view-specific graph denoted as $G^{v}=\{U \cup S^{v}, E^{v}\},v\in\{1,...,m\}$. Here $S^v$ denotes some view-specific nodes. For example, if we use the user-app graph, $S^v$ should be the app set. If we use the relation graph, $S^v$ may be empty. And if we use the attribute graph, $S^v$ may be the attribute sets. 

\begin{table}[htb]
\centering\caption{ Notations and Explanations. }
\label{tab:notation}
\begin{tabular}{|cc|}
\hline
Notation &  Explanation  \\
\hline
$U=U_L \cup U_{ul}$ & User Set \\
\hline
$m$ & Number of views \\
\hline
$n_v$  & Number of nodes in the $v$-th view-specific graph \\
\hline
$y_u$  & the label of the user $u$ \\
\hline
$\mathcal{N}_u^v$  & the neighbors of user $u$ in the $v$-th view-specific graph \\
\hline
$L$  & the number of layers for the view-specific MLP \\
\hline
\end{tabular}
\end{table}

\subsection{\methodshort}
\begin{figure*}[htb]
\centering
\includegraphics[width=0.9\textwidth]{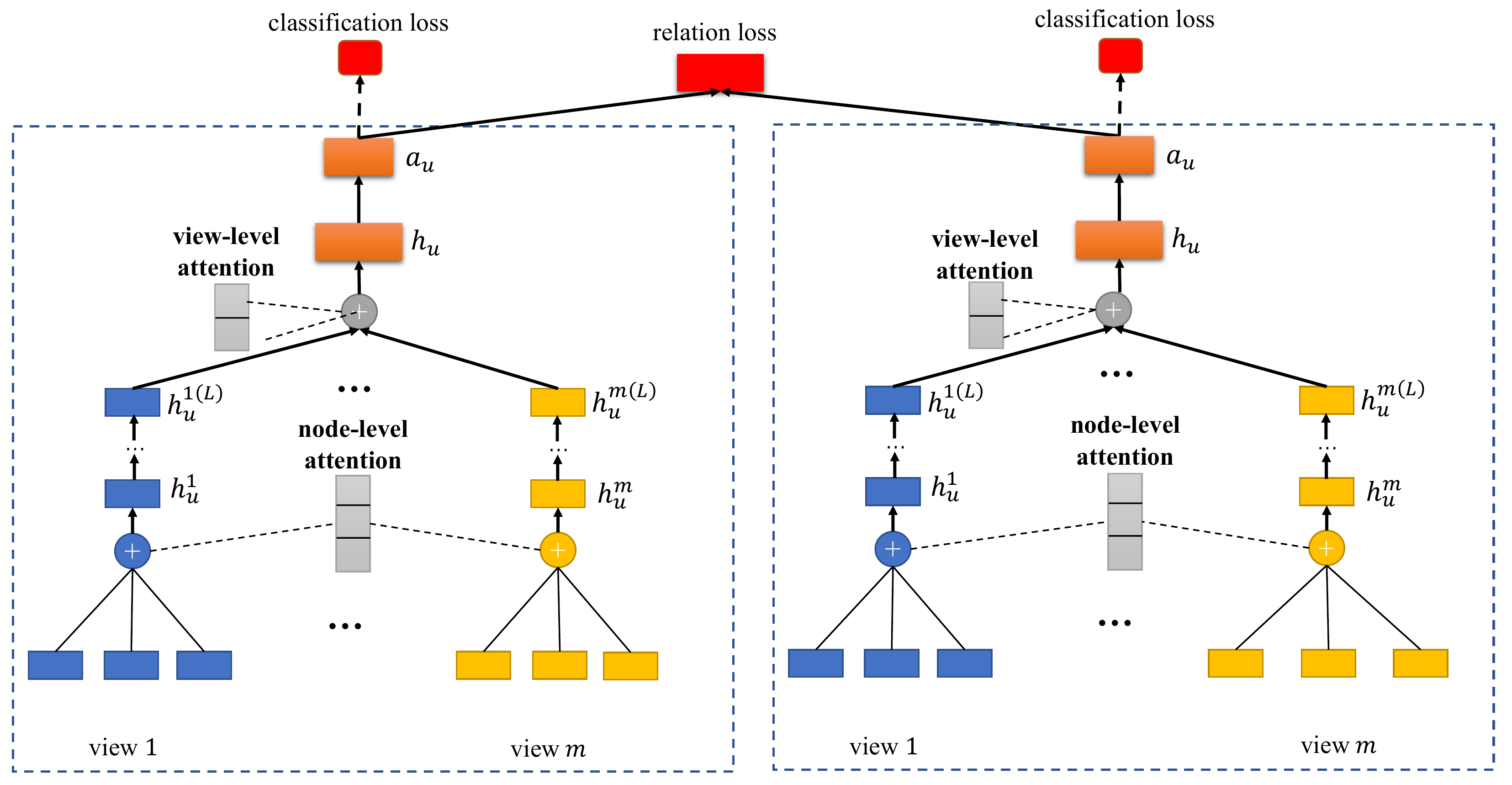}
\caption{The framework of the proposed model \methodshort. }
\label{pic:model_framework}
\end{figure*}

\subsubsection{Overall Architecture}
We introduce how the proposed model \methodshort{} utilizes multiview data for fraud detection here. The model has three key questions to be answered: (1) In each view-specific graph, how to assemble the user's neighbors? (2) How to assemble multiview data to obtain user embedding? (3) How to model the labeled and unlabled user simultaneously? 

We design a hierarchical attention structure in graph neural network: from node-level attention to view-level attention to integrate multiple neighborhoods and different views. The framework of the whole model can be shown in Figure \ref{pic:model_framework}. Firstly, we propose a node-level attention to learn the weight of neighbors for users in each view and accordingly aggregate the neighbors to obtain the low-level view-specific user embeding. Secondly, multiview data characterizes different facets of user information and thus it is essential to integrate multiview data. However, multiview raw data usually have different statistical properites \cite{b30}, which makes it difficult to assmeble multiview data in the low-level space. Since multiview data  describe the same thing, they should show large semantic similarity. Meanwhile, the representations in the high-level space are more close to the semantic. Considering this, we use seperate models to project low-level view-specific user embedding into the high-level space and then integrate view-specific embedding to obtain the joint embedding. Specifically, we propose the view-level attention here to tell the difference of different views and get the optimal combination of view-specific user embedding for the task. Finally, with the combined embedding, we design the supervised classification loss and unsupervised graph-based loss to fully utilize the labeled and unlabeled data. 

\subsubsection{Node-level Attention}
As explained before, we first model each view of graph separately. To model the graph structure, we obtain the user's embedding by ensembling his neighbors' embedding. But we notice that in each view, the neighbors of each user play different roles and show different importance for the specific task. Therefore, we introduce node-level attention here, which can learn the importance of each node to the final task and find a optimal way to aggragate the representations of the neighbors to form the view-specific user embedding. 

Suppose that the node pair $(u,i)$ is connected by an edge with weight of $w_{ui}^v$ in the $v$-th graph, where $u$ represents a user and $i$ represents the user's neighbor. Let $M^{v}\in R^{|n_v|\cdot d}$ denote the node embedding matrix of the $v$-th view, where $d$ denotes the dimension of the node embedding. We use $M^{v}_{i.}$ to denote the embedding look-up operation for the node $i$. Then the weighted embedding of node $i$ can be calculated as $w_{ui} \cdot M^{v}_{i.}$, which we simply denote as $e^v_{ui}$. With $e^v_{ui}$ between each node pair, we hope to inject the structural information into the model and aggregate the neighbors' embedding to obtain the user's embedding. However, different nodes contribute unequally for user's embedding and the final task. Therefore, to learn the importance automatically, we propose a node-level attention mechanism. In detail, let $H^v$ be a learnable attention matrix parameter. Then the importance $\alpha_{ui}^v$ of view-specific node pair $(u,i)$ can be defined as follows:
\begin{equation}
\alpha_{ui}^v = \frac{exp(e_{ui}^v\cdot H_{ui}^v)}{\sum_{k \in \mathcal{N}_u^v}{exp(e_{uk}^v\cdot H_{uk}^v)}}. \nonumber
\end{equation}

Once obtained $\alpha_{ui}^v$, the normalized importance is used to compute a linear combination of the neighbors' embedding to obtain the low-level view-specific user embedding:
\begin{equation}
h_{u}^{v} = \sum_{k \in \mathcal{N}_u^{(A)}}{\alpha_{uk}e_{uk}}, 
\label{eqn:node_aggr}
\end{equation}

Similarily, we can obtain the user embedding of other views in the similar way.

\subsubsection{View-level Attention}
From Eq. \ref{eqn:node_aggr}, we can obtain the view-specific low-level embedding for each user. To learn more comprehensive user embedding, we should fuse multiple views of information. As we state before, low-level representations of multiview data lie in heterogeneous domains, which makes it difficult to capture the multiview correlations in low-level space. To address the problem, we use separate multi-layer perceptrons (MLP) to project low-level view-specific user embedding into the high-level space first and then integrate multiview data. In detail, the representations of the $l$-layer are defined as: 
\begin{equation}
h_{u}^{v(l)} = Relu(h_{u}^{v(l-1)}W_l^{v}+b_l^{v}), v\in\{2,...,m\} ,  
\label{eqn:mlp}
\end{equation}
where $h_{u}^{v(1)}=h_{u}^{v}$.

Through several layers of non-linear functions to map the raw data into the high-level semantic space, it is easier to correlate multiview data \cite{b30}. Now, different views contain different aspect of semantic and each view-specific user embedding only reflect one aspect of information. Meanwhile, considering different aspects contribute differently to the final tasks, we propose a view-level attention mechnism to automatically learn the importance of different views and accordingly integrate multiview data. 

We first introduce a view preference vector $\phi_u^v$ for each user to guide the view-level attention mechanism. The vector is randomly initialized and is learnt during the training process. With the view preference vector $\phi_u^v$, the importance of each view can be calculated as: 
\begin{equation}
\alpha_u^v = \frac{exp(h_{u}^{v(L)} \cdot \phi_u^v)}{\sum\limits_{k \in \{1,...,m\}}{exp(h_{u}^{k(L)}\cdot \phi_u^k)}},v\in\{1,...,m\}, 
\label{eqn:view_preference}
\end{equation}

Here we observe that if the view-specific vector is similar to the preference vector, it will be assigned with larger importance and this view will contribute more to the joint user embedding. 

With the learned view-attention importance, the joint embedding of user $u$ can be obtained by weighted combining view-specific embedding $\{h_{u}^{1(L)},...,h_{u}^{m(L)}\}$:
\begin{equation}
h_{u} = ||_{i=1}^m{(\alpha_u^v \cdot h_{u}^{v(L)})},  
\label{eqn:joint_embedding}
\end{equation}
where $||$ denotes the concatenation operation.

To summarize, the view-level attention mechanism models personalized preference on different views by introducing preference vector for each user and each view. In this way, the joint embedding can naturally distinguish view-specific user embedding and fuse them. 

Finally, we use a one-layer perceptron, which uses the joint user embedding $h_{u}$ as the input to refine the representations. After that,  we can obtain the final high-level embedding for the user $u$, which we denote as $a_{u}$. With $a_{u}$ for each user, we can define the task-specific loss function, which we leave in the next section. 

\subsubsection{Loss function and Optimization}
For the labeled users, we use softmax on the representations of the embedding layer to get the classification result. Thus we can define the classification loss: 
\begin{equation}
\mathcal{L}_{sup} = -\frac{1}{|U_L|}\sum\limits_{u\in U_l}\sum\limits_{i=1}^k{I{(y_u=i)}}\ log\frac{exp(a_u\cdot \theta_i)}{\sum\limits_{j=1}^k{exp(a_u\cdot \theta_j)}},
\label{eqn:sup}
\end{equation}
where $I(\cdot)$ is the indicator function, $k$ is the number of occupations to be predicted and $\theta$ is the parameter of the softmax.

However, since fraud labeling is resource-consuming, only a small portion of the users are labeled, which makes the model difficult to learn a good classifier with such limited labeled data. Although a large number of users are not labeled, we have these users' multiview information. Therefore, we consider utilize unlabeled data to help the model training. But how to select the unlabeled data from the huge pool. Previous works have demonstrated that the fraud always happens within a local graph \cite{b5,b6}. Inspired by this, we use the labeled data as the seeds and then obtain the unlabeled data by extending the one-hop social relations like friends, classmates and workmates from the seed users. In this way, we can utilize such social relations to bridge the labeled and unlabeled data and thus learn the model with the unlabeled data. 

To achieve this, inspired by Deepwalk \cite{b19}, we propose an unsupervised graph-based loss function to refine the whole model. Supposing the relation graph is $G^{(U)}$, we perform random walk to define the neighbors of each vertex. The loss function encourages nearby nodes having similar representaions while makes the representations of disparate nodes distinct:
\begin{eqnarray}
\mathcal{L}_{graph}=\sum\limits_{u \in U}\sum\limits_{v\in \mathcal{N}_u \cup Neg_u}-log(\sigma(a_u^Ta_v)) \nonumber\\ 
-Q\cdot E_{q \sim P_{neg}(u)}log(\sigma(a_u^Ta_q)), 
\label{eqn:unsup}
\end{eqnarray}
where $\mathcal{N}_u$ is the neighbors of the user $u$ and $Neg_u$ is the negative neighbors of the user $u$, $v$ is a node that co-occurs in the $u$'s random walk path, $\sigma$ is the sigmoid function, $P_{neg}(u) \propto d_u^{0.75}$ is the negative sampling distribution and $Q$ is the number of negative samples set to $3$ in our paper. 

In summary, by incorporating the unlabeled data, it can help the model better define the fradulous local structure and the healthy local structure. More importantly, the representations $a_u$ of the unlabeled data we feed into the loss function is generated from the user's multiview information, rather than directly performing embedding look-up. In this way, we not only utilize the social relations of the unlabeled data, but also integrate their content information, which further improves the model's performance. 

Then we combine the supervised classification loss (Eqn. \ref{eqn:sup}) and unsupervised graph loss (Eqn. \ref{eqn:unsup}) to form the final objective function: 
\begin{equation}
\mathcal{L}_{\methodshort{}}=\alpha \cdot \mathcal{L}_{sup}+(1-\alpha) \cdot\mathcal{L}_{graph}+\lambda \mathcal{L}_{reg},
\label{eqn:loss}
\end{equation}
where $\alpha$ is the balancing term between the supervised loss and unsupervised loss and $\mathcal{L}_{reg}$ denotes the $\mathcal{L}$2 regularization of the model parameters. 

We optimize the model using Stochastic Gradient Descent (SGD). The pseudo code to train \methodshort{} is listed in Alg. \ref{alg:train}.
\begin{algorithm}[htb]
\caption{Training Algorithm for \methodshort{}}
\label{alg:train}
\begin{algorithmic}[1]
\REQUIRE The multiview graph: $G^{v}=\{U \cup S^{v}, E^{v}\},v\in\{1,...,m\}$. The balancing weight $\alpha$. The regularizer weight $\lambda$. 
\ENSURE The model parameters $\Theta$
\STATE Randomly initialize the model parameters $\Theta$ and the attention parameter $H^v$ and $\phi^v$. 
\STATE Generating random walk paths according to relation graph $G^{(U)}$ and construct the user paired set S.
\FORALL{$(u,v) \in S$}
\FORALL{$k \in \{1,...,m\}$}{}
\STATE Obtain low-level view-specific user embedding $h_u^k$ and $h_v^k$ by Eq. \ref{eqn:node_aggr}.
\STATE Obtain high-level view-specific user embedding $h_u^{k_L}$ and $h_v^{k_L}$ by Eq. \ref{eqn:mlp}
\STATE Obtain view preference vector $a_u^k$ and $a_v^k$ by Eq. \ref{eqn:view_preference}
\ENDFOR 
\STATE Obtain $a_u$ and $a_v$ by Eq. \ref{eqn:joint_embedding}. 
\STATE Get $\mathcal{L}_{\methodshort{}}$ by Eqn. \ref{eqn:loss}.
\STATE Do backpropagation and update model parameters: $\Theta^{(new)}=\Theta^{(old)}-\lambda \cdot \frac{\partial{\mathcal{L}_{\methodshort{}}}}{\partial{\Theta^{(old)}}}$
\ENDFOR
\RETURN{ $\Theta$}
\end{algorithmic}
\end{algorithm}

\subsubsection{Analysis of the Proposed Model}
Here we give some analysis of the proposed model \methodshort{}:
\begin{itemize}
	\item We propose a general multiview graph model here and the model can deal with various types of nodes and various views of the graphs. Our model can fuse the rich semantics in each view and learn a comprehensive user embedding for user fraud detection. Furthermore, since we model the multiview data in a unified framework and optimize them together, different views can enhance the mutual promotion and mutual upgrade. 
	\item The proposed model potentially has good interpretability for the given task. The node-level attention and the view-level attention are optimized with the classification error and in the end the model will pay more attetion to some meaningful nodes and meaningfull views. In this way, by using the attention term, we can observe which nodes are important for the given task and for a given user, what factors are most influential for the user to be classified as fraud or not. Such results are benefitial to analyze and explain the results and help understand the models. 
	\item The proposed model is an inductive method: Given a new user, if we have its multivew information, we can directly feed the user into the model to decide whether it is fraud or not. And commonly we have the multiview information for almost all the users. In this way, once the model is trained, it can be used to classify all the users on our platform. 
	\item The proposed \methodshort{} is efficient and can be parallelized easily. In the training procedure, we need to go over several iterations. In each iteration, we need to go over the whole edge. For the node-level attention, we will look at the node's neighbors and for the view-level attention, we need to go over the whole views. Therefore, the training complexity of the model is $O(I \cdot |E|\cdot d \cdot m)$, where $I$ is the number of iterations, $d$ is the average degree of a node and $|E|$ is the edge size of the relation graph. Therefore, the complexity of the model is linear to the number of edges. Furthermore, the model can be easily parallelized because the optimization for each edge is independent. Therefore, we can deploy the model into several machines to do optimization. 
\end{itemize}

\section{Experiment}
In this experiment, we conduct experiments on a real-world dataset to answer the following questions:
\begin{itemize}
	\item Can \methodshort{} learn better embeddings for fraud detection compared with existing methods?
	\item Whether the proposed hierarchical attention mechanim can help improve the model's performance and give interpretable results? 
	\item Whether the performance of our model can benefit from the unlabeled data? 
	\item Is \methodshort{} sensitive to the parameters and how the performance will be affected? 
\end{itemize}

\subsection{Dataset}
We use the dataset from \textsc{Alipay}\footnote{https://www.alipay.com}, the biggest third-party online payment platform in China, and through it users are able to do both online and offline payment. \textsc{Alipay} provides a credit service name Huabei to the users. Huabei is like a credit card. The users who subscribe the service will be provided some credits to do online and offline shopping. Then users need to do repayment of Huabei some days later. Based on such a service, we conduct the experiment on two tasks: user default prediction and user attributes prediction. User default prediction helps to predict whether someone has  enough ability to do repayment. In this way the service provider can do some things to prevent the default. And user attribute prediction helps to decide how many credits should be provided for the user.  

We use the following data sources to form the multiview graph. Firstly, we use the user-relation graph. Here, in our problem if two users are labeled as friends, classmates or workmates, there will be an undirected edge with weight $1$ for simplicity between the two users. Secondly, we use the user-app graph. When a user logins an app, there will be an edge connecting with them and the weight denotes the frequency. Thirdly, we use the user-nick graph. The nicks of a user are marked online by other users. The assumption of using nicks is that how people describes other person may help define the user. We use word-cut algorithm to seperate each nick into several words and if a user is marked by a nick, we will connect the user with all the words of the nicks by edges and the edge weight denotes the word frequency. In this way, we can build user-nick bipartite graph. Finally, users usually upload their frequently-used addresses for online merchant. One address is usually a sentence containing the country, province (state), city, street and door information of the user. For most of the users, they have at least one address. Similarily, we split each address into several words. If the user's address contains a word, we set an edge between the user and the word and the edge's weight is corresponding to the word frequency. In this way, each user will link to several words, which forms another bipartite graph. 

We collect about $4$ million users with the known labels. To obtain the unlabeled data, we collect the users which are one-hop friends, classmates and workmates of the labeled users. Then, the total number of users are over $1$ hundred millions. Specifically, we withdraw the addresses which have not been used over the past half of the years. Then about $90$ percent of the users have at least one app, over $95$ percent of the users have the nicks and $80$ percent of the users have at least one address. We use the word-cut package to split each address and the nicks into several words and we withdraw the words whose frequency is below the bottom $10$\%. Totally, the vocabulary size for addresses are $300$ thousand, for nicks are $500$ thousand and for apps are $20$ thousand.

\begin{table*}[htb]
\centering\caption{ User Default Prediction on AUC and KS on \textsc{Alipay}. }
\label{tab:res_default}\small
\begin{tabular}{|c|c|c|c|c|c|c|c|c|}
\hline
Evaluation Metric  & Xgboost & LINE & GCN & GAT & \methodshort$_{sup}$ & \methodshort$_{nd}$ & \methodshort$_{vw}$ & \methodshort  \\
\hline
AUC & 0.753 & 0.771 & 0.780 & 0.784 & 0.786 & 0.798 & 0.801 & \textbf{0.807}  \\
\hline
KS  & 0.370 & 0.397 & 0.415 & 0.424 & 0.427 & 0.442 & 0.448 & \textbf{0.464}   \\
\hline
\end{tabular}
\end{table*}

\begin{figure*}[htb]
\centering
\subfigure[F1-score]{
\includegraphics[width=0.3\textwidth]{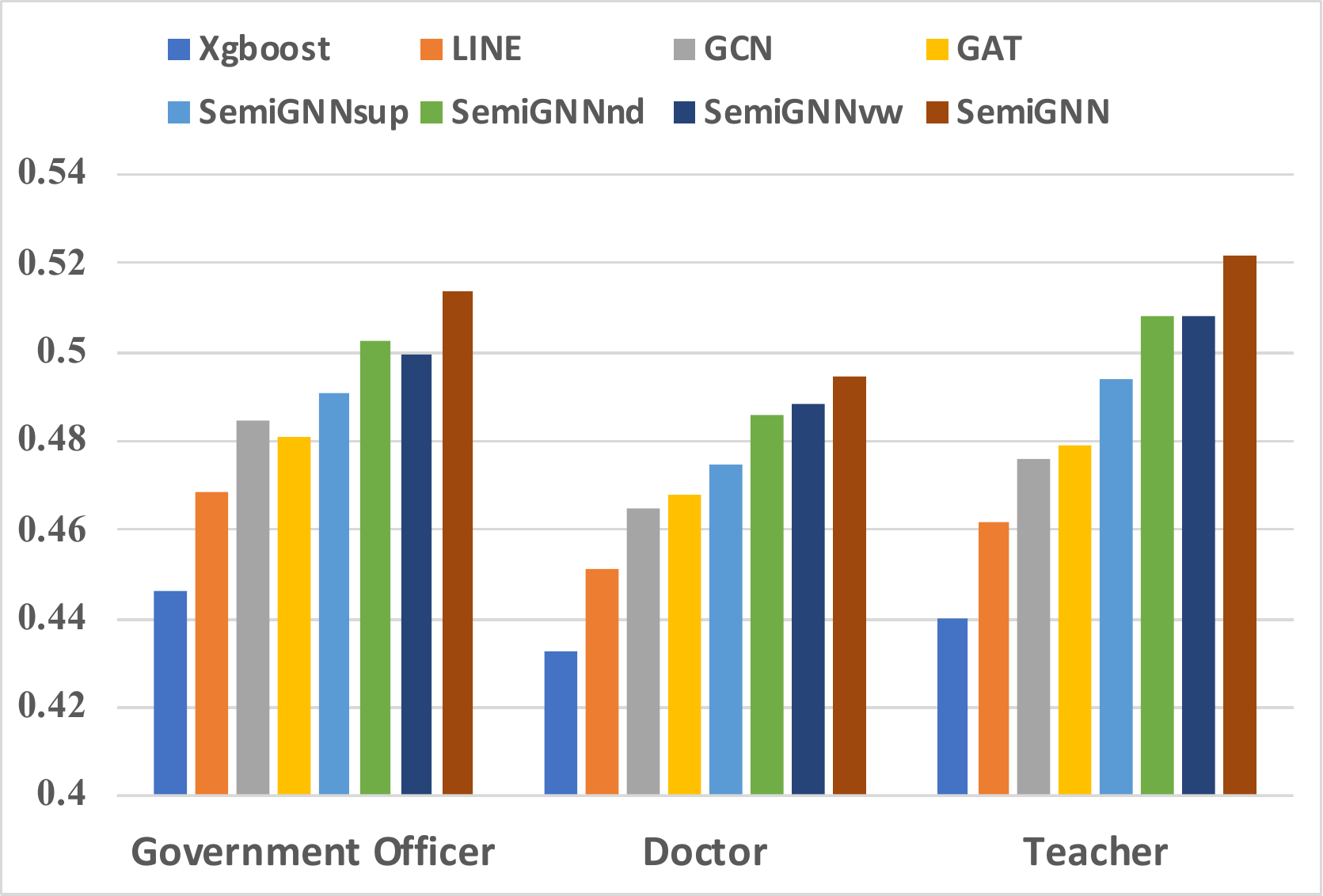} 
\label{pic:f1}}
\subfigure[Precision]{
\includegraphics[width=0.3\textwidth]{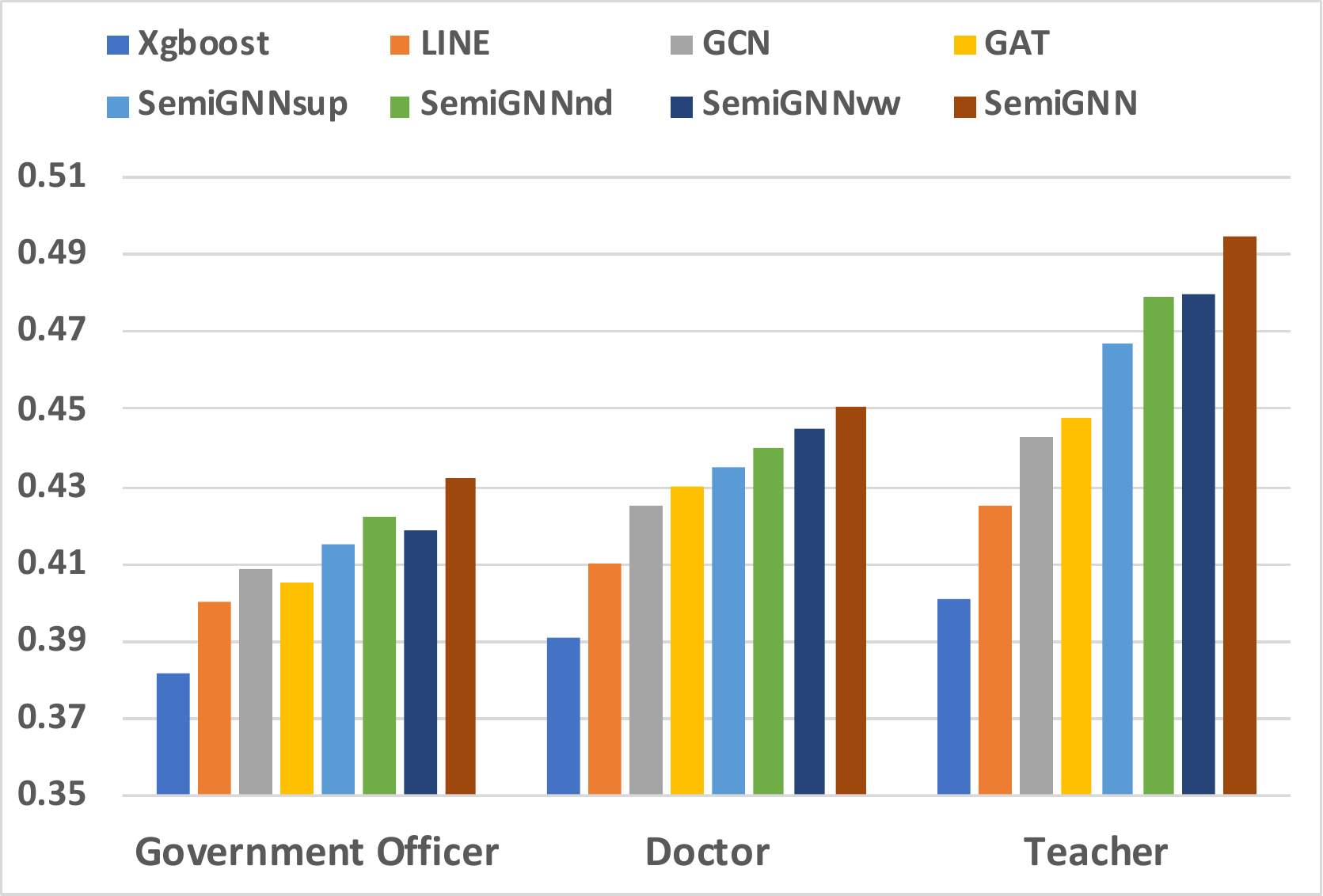}
\label{pic:precision}}
\subfigure[Recall]{
\includegraphics[width=0.3\textwidth]{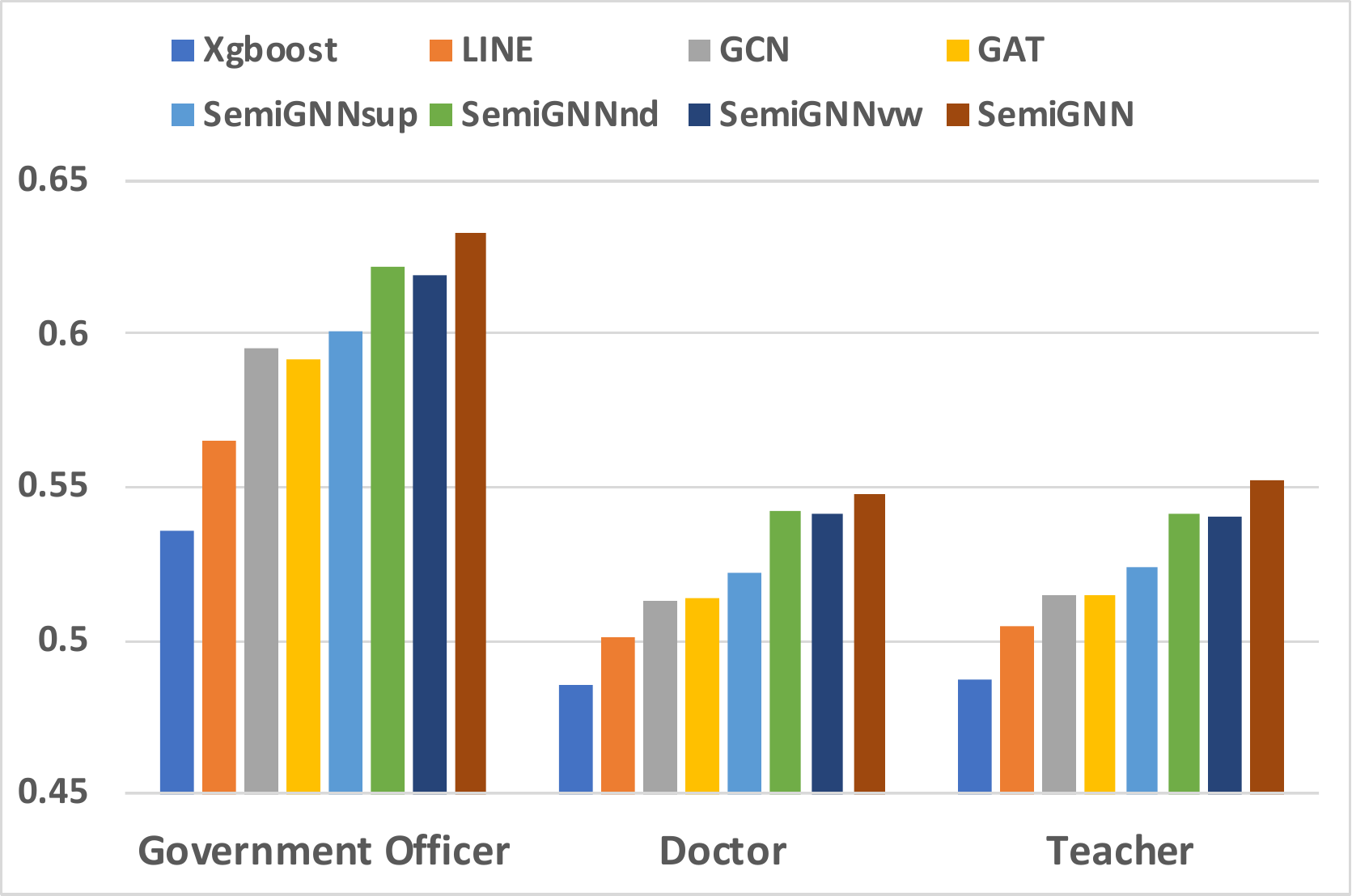}
\label{pic:recall}}
\caption{ Attribute classification results in terms of F1-score, precision and recall.}
\label{pic:result}
\end{figure*}

\subsection{Baseline Methods}
The state-of-the-art methods are introduced here:
\begin{itemize}
	\item Xgboost \cite{b31}: It is a tree-based model, which is very popular and perform well in industry. For each view, we use the dense features as the input to the model.  
	\item LINE \cite{b21}: We first use LINE on the multiview graph to learn the user representations and then use softmax to do classification. Note that DeepWalk, LINE and node2vec are similar in terms of the problem to be solved and the performance. LINE is more scalable and thus we only report the performance of LINE here. 
    \item GCN \cite{b27}: It is a deep graph neural network and each node's embedding is ensembled by its own embedding and the neighbors embedding. Each labeled user is defined as a sample. The node attributes are defined as the average of the pretrained address, nick and app embedding. The graph is defined as the relational graph between the labeled data.
	\item GAT \cite{b32}: It is an improved version of GCN method, which uses the attention mechanism to aggregate the neighborhoods. 
	\item \methodshort$_{sup}$: It is a reduced version of our proposed method \methodshort{}, which only utilizes the labeled data.
	\item \methodshort$_{nd}$: It is a reduced version of \methodshort{} which removes node-level attention.
	\item \methodshort$_{vw}$: It is a reduced version of \methodshort{} which removes view-level attention.
\end{itemize}

\subsection{Implementation Details}
In our model \methodshort{}, we randomly initialize parameters and optimize the model with Adam \cite{b14}. The learning rate is set as $0.002$ and learning rate decay is set as $0.95$. The batch size is set as $128$. We learn the model for $3$ epochs and repeat the experiments for $3$ times and report the averaged results. For the random walk part, each node is sampled $5$ times with the walk length of $10$. The window size is set to $3$. For the deep models, the initial node embedding is set to $128$. After that, a two-layer perceptron with $64$-$32$ units is set to learn the view-specific embeddings for the users. Then another one-layer perceptron with $32$ units is set to learn the view-integrated embeddings. For Xgboost, we use $500$ trees. For GAT, we use three layers with $128$-$64$-$32$ units. For LINE,  we use LINE$_{1st+2nd}$ with the default parameter settings.

\subsection{Quantitative Results}
\subsubsection{User Default Prediction}
In this task, we split the labeled data into three parts: $50$\% for training, $30$\% for test and $20$\% for validation. Among our dataset, $5$\% of the labeled data are labeled as default and the rest are labeled as non-default. Our model aims to predict the default labels directly. Commonly, we use AUC as the evaluation metric. Specifically, financial scenario also concerns about the KS \cite{b15}, which is a metric to measure the risk differentiation of the model. The result is shown in Table \ref
{tab:res_default}.

From Table \ref{tab:res_default}, we have the following observations and analysis: 
\begin{itemize}
	\item We find that our method \methodshort{} and its variants perform better than other methods. It demonstrates the superiority of the proposed model. 
	\item The result that \methodshort{} outperforms \methodshort$_{nd}$ and \methodshort$_{vw}$ demonstrate that the proposed node-level and view-level attention are both essential and help improve the model's performance. 
    \item The result that \methodshort$_{sup}$ outperforms GAT and GCN demonstrates our assumption that different neighbors and diffrent views contribute differently to the target task. Thus, the attention mechanism can better capture their correlations. Meanwhile, by incorporating the attention mechanism, the model can also provide interpretable results, which we will describe in detail in the next section. 
    \item We find that there is a relatively great increase from \methodshort$_{sup}$ to \methodshort{}. It demonstrates that although unlabeled data do not have labels, their social relations with labeled data and their multiview contents  can still provide very valuable information.
	\item The performance of LINE is much worse than other graph-based methods. It demonstrates the importance of collecting information from connected nodes. 
	\item The performance of Xgboost is poor. It demonstrates that the dense features cannot encode enough information. By formulate the problem as a multiview graph and encode each attribute into a vectorized embedding can preserve much more information into the representations, which fasciliates the model learning. 
\end{itemize}

\subsubsection{User Attribute Prediction}
User attributes are often very important for fraud detection because the attributes help to define a person. Specifically, the user's occupation is a very critical attribute to reflect a person's economic and education condition. Therefore, the occupation prediction can help the downstream fraud detection. Here, we conduct the experiment to predict the user's occupation. We first report the results on the common classification metrics F1-score, precision and recall in Figure \ref{pic:result}. Since the occupation prediction of our dataset is mainly used for financial risk control, we also focus on the prediction precision of the top-ranked results, which is shown in Table \ref{tab:top1}.

\begin{table}[htb]
\centering\caption{ Occupation Classification in terms of Top$1$\% Precision on \textsc{Alipay}. }
\label{tab:top1}\small
\begin{tabular}{|c|c|c|c|c|c|c|}
\hline
Method  & Government Officer & Doctor & Teacher  \\
\hline
Xgboost & $0.491$ & $0.575$ & $0.568$  \\
\hline
LINE & $0.539$ & $0.621$ &  $0.609$\\
\hline
GCN & $0.572$ & $0.647$ & $0.637$ \\
\hline
GAT & $0.565$ & $0.649$ & $0.638$ \\
\hline
\methodshort$_{sup}$ & $0.575$ & $0.668$ & $0.659$ \\
\hline
\methodshort$_{nd}$ & $0.597$ & $0.682$ & $0.672$ \\
\hline
\methodshort$_{vw}$ & $0.595$ & $0.679$ & $0.671$ \\
\hline
\methodshort & $\mathbf{0.608}$ &  $\mathbf{0.695}$ & $\mathbf{0.688}$ \\
\hline
\end{tabular}
\end{table}

From Table \ref{tab:top1} and Figure \ref{pic:result}, we find that our method \methodshort{} and its variants still perform better than other methods on the three attributes. It demonstrates the superiority of the proposed model. Other observations are similar with those of the previous section, which we will not repeat any more. 

\subsubsection{The effect of different views}

In this section, we aim to test the effect of different views on two tasks. We use \methodshort{$_x$} to denote the proposed method which only uses the features of $x$ as the input graph. The result is shown in Table \ref{tab:views}.

\begin{table}[ht]
\centering\caption{ KS of the methods when using different views of data. }
\label{tab:views}
\begin{tabular}{|c|c|c|}
\hline
Method  & Default Prediction & Attribute Prediction (Doctor)  \\
\hline
\methodshort$_{app}$ & $0.357$ & $0.554$  \\
\hline
\methodshort$_{addr}$ & $0.303$ & $0.638$  \\
\hline
\methodshort$_{nick}$ & $0.22$ & $0.615$  \\
\hline
\methodshort$_{social}$ & $0.285$ & $0.604$  \\
\hline
\methodshort & $\mathbf{0.464}$ &  $\mathbf{0.729}$  \\
\hline
\end{tabular}
\end{table}

From Table \ref{tab:views}, we have the following observations:
\begin{itemize}
	\item In both tasks, the performance will drop a lot if we only utilize one view of graph compared with the performance of \methodshort{}. It demonstrates that it is very essential to integrate multiple views to get a more comprehensive result. 
	\item In both tasks, we find that social relations are effective. It demonstrates that similar people will gather together and it is essential to utilize the graph model to model such a structure. 
	\item In the task of default prediction, we find that app and address are important. The reason of app achieving good performance is that default people will usually use some apps to borrow the money and spend the money, which greatly increases the loan of the people and as a result they cannot repay the credits in time. The reason that addresses can achieve a good performance is that users often set the frequent delivery address as the home address or work address. These two addresses can indirectly reflect the people's financial condition.  
	\item In the task of occupation attribute prediction, we find that address is the most important features compared with other features. The reason is that in most cases people will use their work address as the delivery address. And the work address can always show the occupation of the users. We find that the app features perform worse in this task. We think the reason is that the apps are nosiy for occupation prediction. Very few apps are specifically designed for specific occupations. 
\end{itemize}

\subsection{Interpretable Results}
We report the interpretable results in this section. In our model, for each user, since we use the node-level attention mechanism to aggregate neighbors' representations, the value of the attention term can been seen as the importance of the neighbor to the final task. Therefore, we aggregate the importance of each node to obtain the node's global importance to the final task. Due to the space limit, we give the top $15$ important apps for default prediction. We also give top $15$ addresses and nicks for attribute prediction on doctor. The reason why we report different features for different tasks is that these features are more useful for a specific task, as shown in the previous section. The result is shown in Table \ref{tab:importance}.

\newcommand{\tabincell}[2]{\begin{tabular}{@{}#1@{}}#2\end{tabular}}
\begin{table*}[htb]
\centering\caption{ Most $15$ important words for user default prediction and user attribute prediction. Actually, the original languages for these words are all Chinese. We translate them into English and report both English and Chinese here. Specifically, for the apps, we not only report the English name and Chinese name, but also report its category for a better understanding. }
\scriptsize
\label{tab:importance}
\begin{tabular}{|c|c||c|c|}
\hline
\multirow{2}{*}{Rank} &
 \multicolumn{1}{c||}{User Default Prediction} &  \multicolumn{2}{c|}{User Attribute Prediction (Doctor)} \\
\cline{2-4} & App & Nick  & Address\\
\hline
1 & \textbf{game}-jjhgame (集结号捕鱼)  &  Head Nurse  (护士长) &  Maternity Hospital (妇产医院)   \\
\hline
2 & \textbf{p2p}-crfchina \ (信而富) & Dean \ (院长) & Pet Hospital \ (宠物医院)  \\
\hline
3 & \textbf{p2p}-iqianjin \ (钱站)  & Clinic \ (诊所) & Dentistry \ (牙科)   \\
\hline
4 & \textbf{game}-templerun2  \ (神庙逃亡2)  & Doctor \ (医生) & Outpatient Department \ (门诊部)  \\
\hline
5 & \textbf{financial}-eastmoney  \ (东方财富) & Hospital of Chinese Medicine \ (中医院) & Clinic \ (诊所)  \\
\hline
6 & \textbf{p2p}-xiangqd  \ (向钱贷)  & Patient \ (病人) & Physical Examination \ (体检) \\
\hline
7 & \textbf{p2p}-niwodai  \ (你我贷借款)  &  Nurse \ (护士) & Stomatology Department \ (口腔科)  \\
\hline
8 &  \textbf{p2p}-360jie \ (360借条) & Beauty \ (美容院) & Traditional CM Department \ (传统中医科) \\
\hline
9 & \textbf{shopping}-aldb  \ (魔buy商城)  & Attending Doctor \ (主治医师) & Hospital \ (医院) \\
\hline
10 & \textbf{p2p}-jiedaibao \ (借贷宝)  & Dentist \ (牙医) & Gynecology \ (妇科) \\
\hline
11 & \textbf{game}-lua850  \ (850棋牌李逵捕鱼)  & Health-center \ (体检中心) & Rehabilitation Department \ (康复科) \\
\hline
12 & \textbf{entertainment}-cashcomic \ (惠动漫) & Cosmetologist \ (美容师) & Nursing Department \ (护理部) \\
\hline
13 & \textbf{shopping}-globalscanner \ (环球捕手)  & Wardmate \ (病友) & Health Department \ (卫生部) \\
\hline
14 & \textbf{social}-my \ (秒缘) & Radiology \ (放射科) & Pediatric Department \ (儿科)  \\
\hline
15 & \textbf{p2p}-daima360 \ (贷嘛)  & Gynecology \ (妇科) &  Obstetrics Department \ (产科) \\
\hline
\end{tabular}
\end{table*}

\begin{figure*}[htb]
\centering
\subfigure[The Dimension of Final Embedding]{
\includegraphics[width=0.3\textwidth]{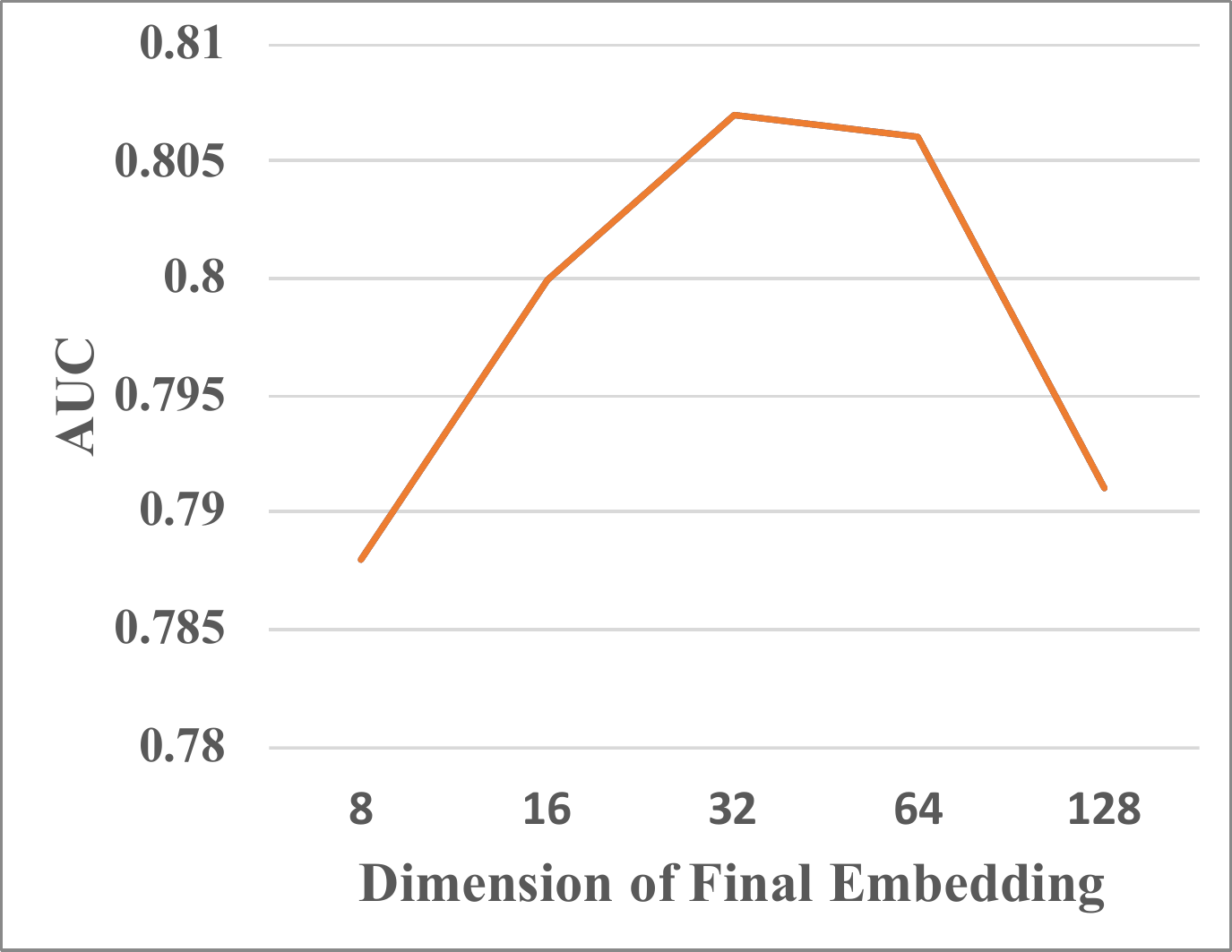} 
\label{pic:dim_final}}
\subfigure[The Dimension of Initial Node Embedding]{
\includegraphics[width=0.3\textwidth]{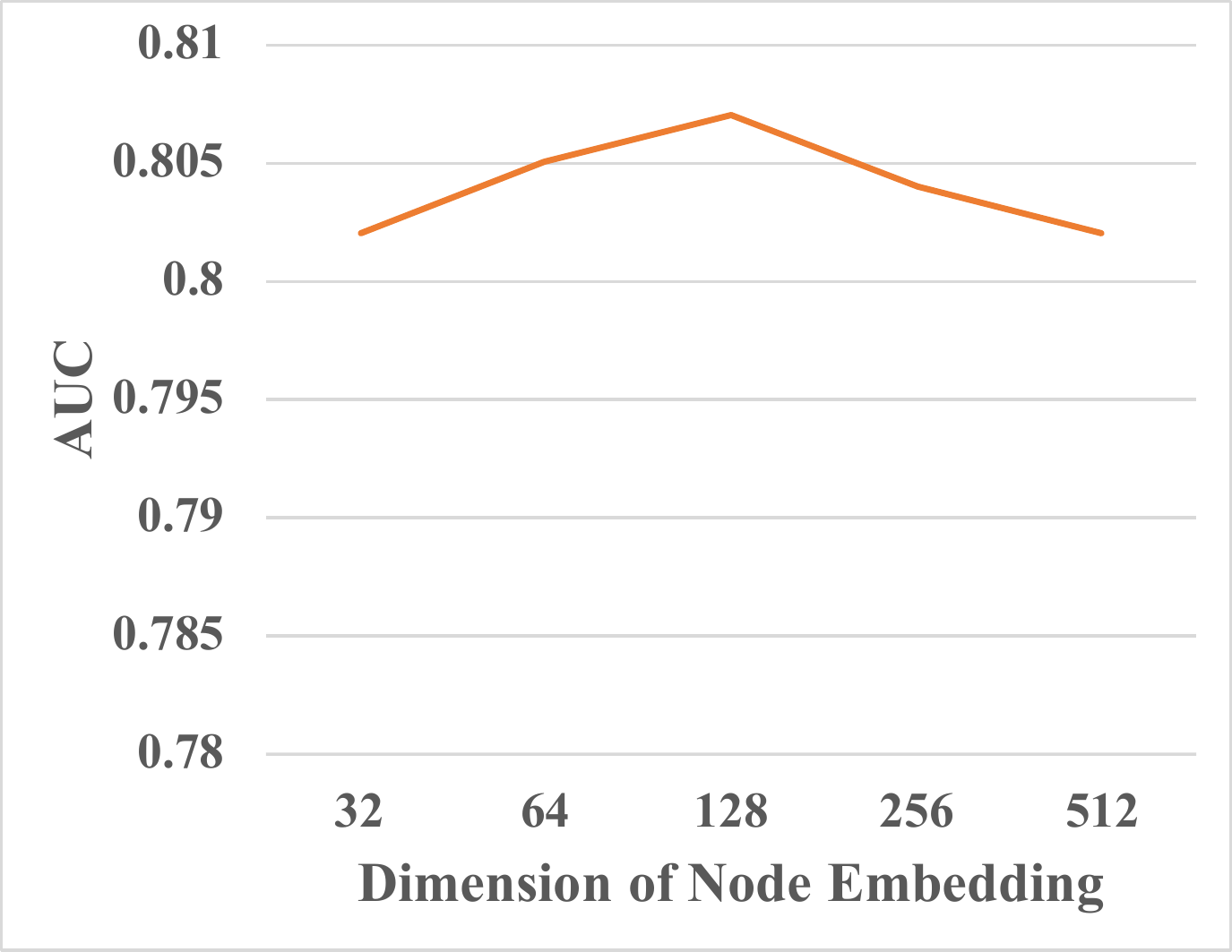}
\label{pic:dim_node}}
\subfigure[The Value of $\alpha$]{
\includegraphics[width=0.3\textwidth]{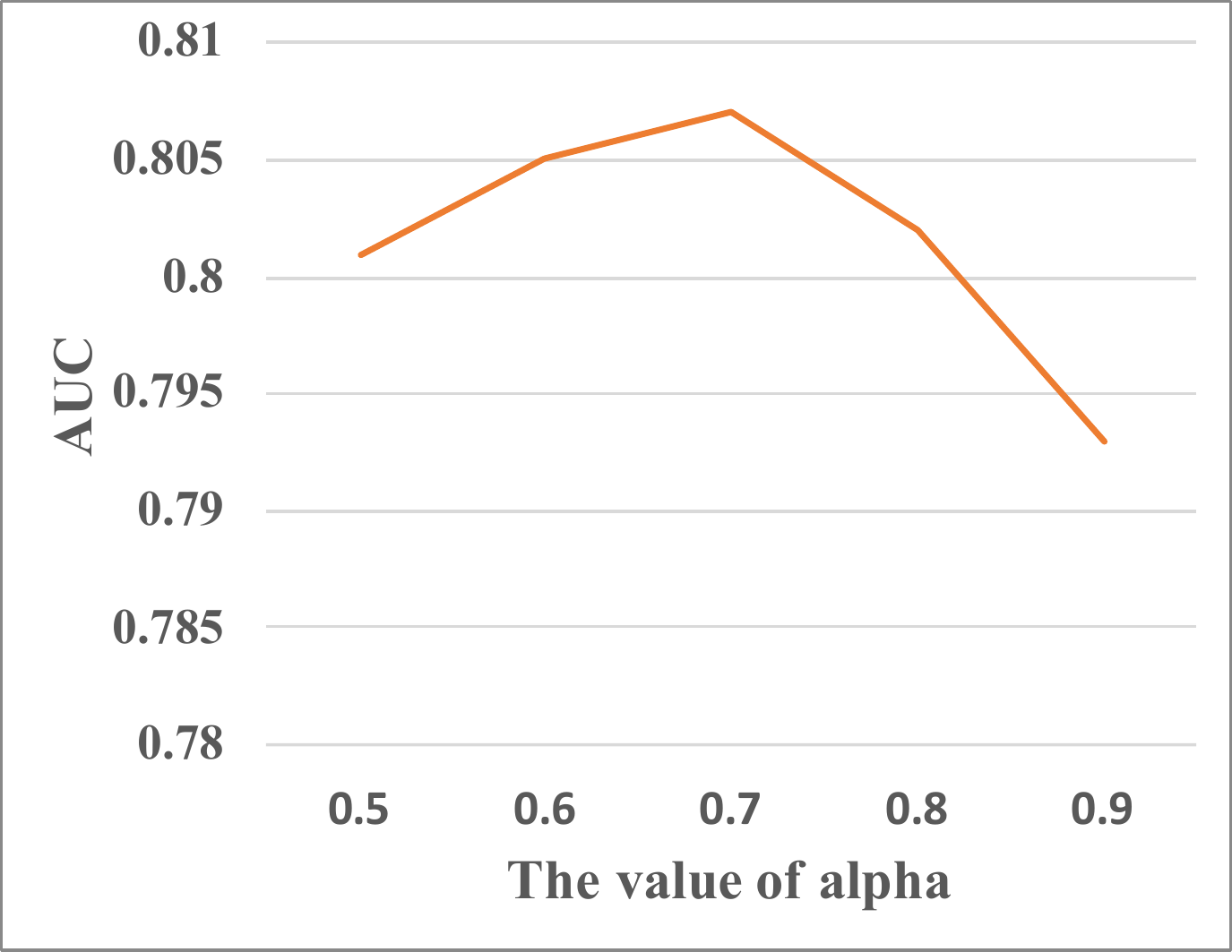}
\label{pic:alpha}}
\caption{ Parameter Sensitivity on dimension of final embedding, dimension of node embedding and the value of $\alpha$ on the task of default prediction.}
\label{pic:sensitivity}
\end{figure*}

From Table \ref{tab:importance}, we find that almost all the features are highly related to the corresponding tasks. Specifically, we also get the following interesting findings:
\begin{itemize} 
    \item We observe that most of the important apps for default prediction are financial apps, especially the p2p apps. It is reasonable because if people frequently use the p2p apps, their loan will be larger which may lead to the default finally. We also find that the game apps and shopping apps are also important. The reason is that users may spend a lot of money on these apps. In this way, they may not have enough money to repay the credit, which leads to the default. 
    \item We observe that in user attribute prediction, most of the important nicks are titles related to the doctor. Such a result is very reasonable in China because people often mark other peopole's name with their occupations, like Doctor Wang and Nurse Peter. Therefore, we can find useful information in the nicks. When we look at the important addresses for doctor, we find that our method is able to find many departments of the hospital. They are very typical features for doctors, which also demonstrates a great interpretability of our method. 
\end{itemize}

\subsection{Parameter Sensitivity}
In this section, we investigate the parameter sensitivity. We change the value of one parameter and fix the values of other parameters. Then we report the results of AUC on \textsc{Alipay} dataset with various parameter settings in Figure \ref{pic:sensitivity}. Due to the space of limit, we only report the results on the task of default prediction.

\begin{itemize}
	\item Dimension of the final embedding: We first test the effect of the dimension of the final embedding on the classification performance. The result is shown in Figure \ref{pic:dim_final}. We can see that with the growth of the embedding dimension, the AUC raises and then drops. The reason is that  a suitable dimension is needed to encode enough information but a larger dimension will lead to overfitting and redundancy. 
	\item Dimension of the initial node embedding: Then we report the effect of the dimension of the initial node embedding in Figure \ref{pic:dim_node}. We can see that generally there is little change in terms of AUC when we change the node embedding dimension. The reason is that these dimensions are enough to encode useful information. 
	\item The value of $\alpha$: We change the value of $\alpha$ and report the corresponding performance in Figure \ref{pic:alpha}. We find that a better performance will achieve when we pay more attention to the labeled data because labeled data can provide discriminatve information. But we also observe that ignoring the unsupervised information also makes the performance worse. The reason is that the unsupervised information can provide structural information, which also benefits the model learning. 
\end{itemize}

\section{Conclusion}
In this paper, we propose a semi-supvervised graph attentive network model for fraud detection. Our model links the labeled and unlabeled data via their social relations. And we learn a classifier which on the one hand is consistant with the labels of the labeled data by proposing the classification loss, and on the other hand makes the classification results for similar vertexes similar by proposing the graph-based loss. Specifically, we propose a hierarchical attention mechanism to better mine the multiview graph. The node-level attention is able to better correlate neighbors and the view-level attention can better integrate different views. Experimentally, our method achieves better results compared with baseline methods. And our method can tell important factors for a specific task. 

The future work may focus on differentiating different social relations to further improve the model. And we can extend the model to more fraud detection applications.  

\section{Acknowledgement}
We would like to thank the support of many colleagues from AI lab and the Financial Risk Management in Ant Financial Services Group. Thanks for the support of the China Postdoctoral Science Foundation.

\vspace{12pt}
\end{CJK}

\begin{thebibliography}{}

\bibitem[\protect\citeauthoryear{Bernstein}{1960}]{bernstein1960language}
Basil Bernstein.
\newblock Language and social class.
\newblock {\em The British journal of sociology}, 11(3):271--276, 1960.

\bibitem[\protect\citeauthoryear{Bradley \bgroup \em et al.\egroup
  }{2000}]{bradley2000case}
Keith Bradley, Rachael Rafter, and Barry Smyth.
\newblock Case-based user profiling for content personalisation.
\newblock In {\em International Conference on Adaptive Hypermedia and Adaptive
  Web-Based Systems}, pages 62--72. Springer, 2000.

\bibitem[\protect\citeauthoryear{Cao \bgroup \em et al.\egroup
  }{2015}]{Cao2015GraRep}
Shaosheng Cao, Wei Lu, and Qiongkai Xu.
\newblock Grarep: Learning graph representations with global structural
  information.
\newblock In {\em CIKM}, pages 891--900, 2015.

\bibitem[\protect\citeauthoryear{Caruso \bgroup \em et al.\egroup
  }{2013}]{caruso2013user}
Mario Caruso, Massimo Mecella, Roberto Baldoni, Leonardo Querzoni, and Adriano
  Cerocchi.
\newblock User profiling and micro-accounting for smart energy management.
\newblock In {\em Proceedings of the 11th ACM Conference on Embedded Networked
  Sensor Systems}, page~42. ACM, 2013.

\bibitem[\protect\citeauthoryear{Chang \bgroup \em et al.\egroup
  }{2015}]{chang2015heterogeneous}
Shiyu Chang, Wei Han, Jiliang Tang, Guo-Jun Qi, Charu~C Aggarwal, and Thomas~S
  Huang.
\newblock Heterogeneous network embedding via deep architectures.
\newblock In {\em SIGKDD}, pages 119--128. ACM, 2015.

\bibitem[\protect\citeauthoryear{Chen and Guestrin}{2016}]{chen2016xgboost}
Tianqi Chen and Carlos Guestrin.
\newblock Xgboost: A scalable tree boosting system.
\newblock In {\em Proceedings of the 22nd acm sigkdd international conference
  on knowledge discovery and data mining}, pages 785--794. ACM, 2016.

\bibitem[\protect\citeauthoryear{Dong \bgroup \em et al.\egroup
  }{2017}]{dong2017metapath2vec}
Yuxiao Dong, Nitesh~V Chawla, and Ananthram Swami.
\newblock metapath2vec: Scalable representation learning for heterogeneous
  networks.
\newblock In {\em SIGKDD}, pages 135--144. ACM, 2017.

\bibitem[\protect\citeauthoryear{French}{1959}]{french1959can}
Wendell~L French.
\newblock Can a man's occupation be predicted?
\newblock {\em Journal of Counseling Psychology}, 6(2):95, 1959.

\bibitem[\protect\citeauthoryear{Grover and
  Leskovec}{2016}]{Grover2016node2vec}
Aditya Grover and Jure Leskovec.
\newblock node2vec:scalable feature learning for networks.
\newblock In {\em SIGKDD}, pages 855--864, 2016.

\bibitem[\protect\citeauthoryear{Kipf and Welling}{2016}]{kipf2016semi}
Thomas~N Kipf and Max Welling.
\newblock Semi-supervised classification with graph convolutional networks.
\newblock {\em arXiv preprint arXiv:1609.02907}, 2016.

\bibitem[\protect\citeauthoryear{Kosinski \bgroup \em et al.\egroup
  }{2013}]{kosinski2013private}
Michal Kosinski, David Stillwell, and Thore Graepel.
\newblock Private traits and attributes are predictable from digital records of
  human behavior.
\newblock {\em Proceedings of the National Academy of Sciences}, page
  201218772, 2013.

\bibitem[\protect\citeauthoryear{Labov}{2006}]{labov2006social}
William Labov.
\newblock {\em The social stratification of English in New York city}.
\newblock Cambridge University Press, 2006.

\bibitem[\protect\citeauthoryear{Li \bgroup \em et al.\egroup
  }{2014}]{li2014weakly}
Jiwei Li, Alan Ritter, and Eduard Hovy.
\newblock Weakly supervised user profile extraction from twitter.
\newblock In {\em Proceedings of the 52nd Annual Meeting of the Association for
  Computational Linguistics (Volume 1: Long Papers)}, volume~1, pages 165--174,
  2014.

\bibitem[\protect\citeauthoryear{Lin \bgroup \em et al.\egroup
  }{2014}]{lin2014new}
Jovian Lin, Kazunari Sugiyama, Min-Yen Kan, and Tat-Seng Chua.
\newblock New and improved: modeling versions to improve app recommendation.
\newblock In {\em Proceedings of the 37th international ACM SIGIR conference on
  Research \& development in information retrieval}, pages 647--656. ACM, 2014.

\bibitem[\protect\citeauthoryear{Lindquist \bgroup \em et al.\egroup
  }{1997}]{lindquist1997influence}
Thalina~L Lindquist, Lawrence~J Beilin, and Matthew~W Knuiman.
\newblock Influence of lifestyle, coping, and job stress on blood pressure in
  men and women.
\newblock {\em Hypertension}, 29(1):1--7, 1997.

\bibitem[\protect\citeauthoryear{Liu \bgroup \em et al.\egroup
  }{2012}]{liu2012robust}
Wei Liu, Jun Wang, and Shih-Fu Chang.
\newblock Robust and scalable graph-based semisupervised learning.
\newblock {\em Proceedings of the IEEE}, 100(9):2624--2638, 2012.

\bibitem[\protect\citeauthoryear{Mezghani \bgroup \em et al.\egroup
  }{2012}]{mezghani2012user}
Manel Mezghani, Corinne~Amel Zayani, Ikram Amous, and Faiez Gargouri.
\newblock A user profile modelling using social annotations: a survey.
\newblock In {\em Proceedings of the 21st International Conference on World
  Wide Web}, pages 969--976. ACM, 2012.

\bibitem[\protect\citeauthoryear{Ngiam \bgroup \em et al.\egroup
  }{2011}]{ngiam2011multimodal}
Jiquan Ngiam, Aditya Khosla, Mingyu Kim, Juhan Nam, Honglak Lee, and Andrew~Y
  Ng.
\newblock Multimodal deep learning.
\newblock In {\em Proceedings of the 28th international conference on machine
  learning (ICML-11)}, pages 689--696, 2011.

\bibitem[\protect\citeauthoryear{Niepert \bgroup \em et al.\egroup
  }{2016}]{niepert2016learning}
Mathias Niepert, Mohamed Ahmed, and Konstantin Kutzkov.
\newblock Learning convolutional neural networks for graphs.
\newblock In {\em International conference on machine learning}, pages
  2014--2023, 2016.

\bibitem[\protect\citeauthoryear{Ou \bgroup \em et al.\egroup
  }{2016}]{Ou2016Asymmetric}
Mingdong Ou, Peng Cui, Jian Pei, Ziwei Zhang, and Wenwu Zhu.
\newblock Asymmetric transitivity preserving graph embedding.
\newblock In {\em SIGKDD}, pages 1105--1114, 2016.

\bibitem[\protect\citeauthoryear{Perozzi and Skiena}{2015}]{perozzi2015exact}
Bryan Perozzi and Steven Skiena.
\newblock Exact age prediction in social networks.
\newblock In {\em Proceedings of the 24th International Conference on World
  Wide Web}, pages 91--92. ACM, 2015.

\bibitem[\protect\citeauthoryear{Perozzi \bgroup \em et al.\egroup
  }{2014}]{perozzi2014deepwalk}
Bryan Perozzi, Rami Al-Rfou, and Steven Skiena.
\newblock Deepwalk: Online learning of social representations.
\newblock In {\em SIGKDD}, pages 701--710. ACM, 2014.

\bibitem[\protect\citeauthoryear{Preo{\c{t}}iuc-Pietro \bgroup \em et
  al.\egroup }{2015}]{preoctiuc2015analysis}
Daniel Preo{\c{t}}iuc-Pietro, Vasileios Lampos, and Nikolaos Aletras.
\newblock An analysis of the user occupational class through twitter content.
\newblock In {\em Proceedings of the 53rd Annual Meeting of the Association for
  Computational Linguistics and the 7th International Joint Conference on
  Natural Language Processing (Volume 1: Long Papers)}, volume~1, pages
  1754--1764, 2015.

\bibitem[\protect\citeauthoryear{Schmidt and
  Strauss}{1975}]{schmidt1975prediction}
Peter Schmidt and Robert~P Strauss.
\newblock The prediction of occupation using multiple logit models.
\newblock {\em International Economic Review}, pages 471--486, 1975.

\bibitem[\protect\citeauthoryear{Strully}{2009}]{strully2009job}
Kate~W Strully.
\newblock Job loss and health in the us labor market.
\newblock {\em Demography}, 46(2):221--246, 2009.

\bibitem[\protect\citeauthoryear{Tan \bgroup \em et al.\egroup
  }{2011}]{tan2011user}
Chenhao Tan, Lillian Lee, Jie Tang, Long Jiang, Ming Zhou, and Ping Li.
\newblock User-level sentiment analysis incorporating social networks.
\newblock In {\em Proceedings of the 17th ACM SIGKDD international conference
  on Knowledge discovery and data mining}, pages 1397--1405. ACM, 2011.

\bibitem[\protect\citeauthoryear{Tang \bgroup \em et al.\egroup
  }{2015}]{tang2015line}
Jian Tang, Meng Qu, Mingzhe Wang, Ming Zhang, Jun Yan, and Qiaozhu Mei.
\newblock Line: Large-scale information network embedding.
\newblock In {\em WWW}, pages 1067--1077, 2015.

\bibitem[\protect\citeauthoryear{Veli{\v{c}}kovi{\'c} \bgroup \em et al.\egroup
  }{2017}]{velivckovic2017graph}
Petar Veli{\v{c}}kovi{\'c}, Guillem Cucurull, Arantxa Casanova, Adriana Romero,
  Pietro Lio, and Yoshua Bengio.
\newblock Graph attention networks.
\newblock {\em arXiv preprint arXiv:1710.10903}, 2017.

\bibitem[\protect\citeauthoryear{Wang \bgroup \em et al.\egroup
  }{2016}]{Wang2016Structural}
Daixin Wang, Peng Cui, and Wenwu Zhu.
\newblock Structural deep network embedding.
\newblock In {\em SIGKDD}, pages 1225--1234, 2016.

\bibitem[\protect\citeauthoryear{Xie \bgroup \em et al.\egroup
  }{2016}]{xie2016learning}
Min Xie, Hongzhi Yin, Hao Wang, Fanjiang Xu, Weitong Chen, and Sen Wang.
\newblock Learning graph-based poi embedding for location-based recommendation.
\newblock In {\em Proceedings of the 25th ACM International on Conference on
  Information and Knowledge Management}, pages 15--24. ACM, 2016.

\end{thebibliography}


\begin{thebibliography}{00}
\small
\bibitem{b1} West, Jarrod, Maumita Bhattacharya, and Rafiqul Islam. "Intelligent financial fraud detection practices: an investigation." International Conference on Security and Privacy in Communication Networks. Springer, Cham, 2014.
\bibitem{b2} Hoogs, Bethany, et al. "A genetic algorithm approach to detecting temporal patterns indicative of financial statement fraud." Intelligent Systems in Accounting, Finance and Management: International Journal 15.1‐2 (2007): 41-56.
\bibitem{b3} Quah, Jon TS, and M. Sriganesh. "l 35.4 (2008): 1721-1732.
\bibitem{b4} Yue, Dianmin, et al. "A review of data mining-based financial fraud detection research." 2007 International Conference on Wireless Communications, Networking and Mobile Computing. Ieee, 2007.
\bibitem{b5} Liu, Ziqi, et al. "Heterogeneous Graph Neural Networks for Malicious Account Detection." Proceedings of the 27th ACM International Conference on Information and Knowledge Management. ACM, 2018.
\bibitem{b6} Hu, Binbin, et al. "Cash-out User Detection based on Attributed Heterogeneous Information Network with a Hierarchical Attention Mechanism." (2019).
\bibitem{b7} Bose, I., and J. Wang. "Data mining for detection of financial statement fraud in Chinese Companies." International joint Conference on e-Commerce, e-Administration, e-Society, and e-Education. International Business Academics Consortium (IBAC) and Knowledge Association of Taiwan (KAT), 2007.
\bibitem{b8} Kirkos, Efstathios, Charalambos Spathis, and Yannis Manolopoulos. "Data mining techniques for the detection of fraudulent financial statements." Expert systems with applications 32.4 (2007): 995-1003.
\bibitem{b9} Ravisankar, Pediredla, et al. "Detection of financial statement fraud and feature selection using data mining techniques." Decision Support Systems 50.2 (2011): 491-500.
\bibitem{b10} Bhattacharyya, Siddhartha, et al. "Data mining for credit card fraud: A comparative study." Decision Support Systems 50.3 (2011): 602-613.
\bibitem{b11} Whitrow, Christopher, et al. "Transaction aggregation as a strategy for credit card fraud detection." Data mining and knowledge discovery 18.1 (2009): 30-55.
\bibitem{b12} Bermúdez, Ll, et al. "A Bayesian dichotomous model with asymmetric link for fraud in insurance." Insurance: Mathematics and Economics 42.2 (2008): 779-786.
\bibitem{b13} Viaene, Stijn, et al. "Strategies for detecting fraudulent claims in the automobile insurance industry." European Journal of Operational Research 176.1 (2007): 565-583.
\bibitem{b14} Kingma, Diederik P., and Jimmy Ba. "Adam: A method for stochastic optimization." arXiv preprint arXiv:1412.6980 (2014).
\bibitem{b15} Friedman, Jerome H. "A recursive partitioning decision rule for nonparametric classification." IEEE Transactions on Computers 4 (1977): 404-408.
%\bibitem{b16} 
%\bibitem{b17} 
%\bibitem{b18} 
\bibitem{b19} SIG-KDD, A. C. M. "DeepWalk: Online Learning of Social Representations." (2014).
\bibitem{b20} Grover, Aditya, and Jure Leskovec. "node2vec: Scalable feature learning for networks." Proceedings of the 22nd ACM SIGKDD international conference on Knowledge discovery and data mining. ACM, 2016.
\bibitem{b21} Tang, Jian, et al. "Line: Large-scale information network embedding." Proceedings of the 24th international conference on world wide web. International World Wide Web Conferences Steering Committee, 2015.
\bibitem{b22} Wang, Daixin, Peng Cui, and Wenwu Zhu. "Structural deep network embedding." Proceedings of the 22nd ACM SIGKDD international conference on Knowledge discovery and data mining. ACM, 2016.
\bibitem{b23} Ou, Mingdong, et al. "Asymmetric transitivity preserving graph embedding." Proceedings of the 22nd ACM SIGKDD international conference on Knowledge discovery and data mining. ACM, 2016.
\bibitem{b24} Cao, Shaosheng, Wei Lu, and Qiongkai Xu. "Grarep: Learning graph representations with global structural information." Proceedings of the 24th ACM international on conference on information and knowledge management. ACM, 2015.
\bibitem{b25} Dong, Yuxiao, Nitesh V. Chawla, and Ananthram Swami. "metapath2vec: Scalable representation learning for heterogeneous networks." Proceedings of the 23rd ACM SIGKDD international conference on knowledge discovery and data mining. ACM, 2017.
\bibitem{b26} Chang, Shiyu, et al. "Heterogeneous network embedding via deep architectures." Proceedings of the 21th ACM SIGKDD International Conference on Knowledge Discovery and Data Mining. ACM, 2015.
\bibitem{b27} Kipf, Thomas N., and Max Welling. "Semi-supervised classification with graph convolutional networks." arXiv preprint arXiv:1609.02907 (2016).
\bibitem{b28} Niepert, Mathias, Mohamed Ahmed, and Konstantin Kutzkov. "Learning convolutional neural networks for graphs." International conference on machine learning. 2016.
\bibitem{b29} Liu, Wei, Jun Wang, and Shih-Fu Chang. "Robust and scalable graph-based semisupervised learning." Proceedings of the IEEE 100.9 (2012): 2624-2638.
\bibitem{b30} Ngiam, Jiquan, et al. "Multimodal deep learning." Proceedings of the 28th international conference on machine learning (ICML-11). 2011.
\bibitem{b31} Chen, Tianqi, and Carlos Guestrin. "Xgboost: A scalable tree boosting system." Proceedings of the 22nd acm sigkdd international conference on knowledge discovery and data mining. ACM, 2016.
\bibitem{b32} Veličković, Petar, et al. "Graph attention networks." arXiv preprint arXiv:1710.10903 (2017).
\end{thebibliography}
\end{document}